\documentclass[prd,10pt,twocolumn]{revtex4}

\usepackage{hyperref}
\usepackage{graphicx}
\usepackage{latexsym}
\usepackage{epsfig}
\usepackage[english]{babel}
\usepackage[T1]{fontenc}
\usepackage[utf8]{inputenc}
\usepackage{amsmath}
\usepackage{bbold}

\usepackage{comment}

\usepackage{xcolor}

\date{\today}

\begin{document}
\title{Stability of Schwarzschild (Anti)de Sitter black holes in Conformal Gravity}
\author{Daniele Lanteri$^{\rm 1,2,3}$, Shen-Song Wan$^{\rm 4}$, Alfredo Iorio$^{\rm 2}$, and Paolo Castorina$^{\rm 1,2,4}$}
\affiliation{
\mbox{${}^1$ Istituto Nazionale di Fisica Nucleare, Sezione di Catania, I-95123 Catania, Italy} \\
\mbox{${}^2$ Institute of Particle and Nuclear Physics, Faculty of Mathematics and Physics, Charles University} \\
\mbox{V Hole\v{s}ovi\v{c}k\'ach 2, 18000 Prague 8, Czech Republic}\\
\mbox{${}^3$ Dipartimento di Fisica e Astronomia, Universit\`a  di Catania, I-95123 Catania, Italy.} \\
\mbox{${}^4$} School of Nuclear Science and Technology, Lanzhou University, 222 South Tianshui Road, Lanzhou 730000, China}
		
\begin{abstract}
We study the thermodynamics of spherically symmetric, neutral and non-rotating black holes in conformal (Weyl) gravity. To this end, we apply different methods: (i) the evaluation of the specific heat; (ii) the study of the entropy concavity; (iii) the geometrical approach to thermodynamics known as \textit{thermodynamic geometry}; (iv) the Poincar\'{e} method that relates equilibrium and out-of-equilibrium thermodynamics. We show that the thermodynamic geometry approach can be applied to conformal gravity too, because all the key thermodynamic variables are insensitive to Weyl scaling. The first two methods, (i) and (ii), indicate that the entropy of a de Sitter black hole is always in the interval $2/3\leq S\leq 1$, whereas thermodynamic geometry suggests that, at $S=1$, there is a second order phase transition to an Anti de Sitter black hole. On the other hand, we obtain from the Poincar\'{e} method (iv) that black holes whose entropy is $S < 4/3$ are stable or in a saddle-point, whereas when $S>4/3$ they are always unstable, hence there is no definite answer on whether such transition occurs.
\end{abstract}



\maketitle

\section{Introduction}
Following Bekenstein's \cite{Bekenstein:1972and1973} and Hawking's \cite{Hawking:1974and1975and1976} pioneering work on black holes (BHs), their thermodynamic properties have attracted an enormous interest, ever since. On the other hand, in recent years, an alternative theory of gravity, known as {\it Conformal Gravity} (CG) (see, e.g., the review \cite{Mannheim:2011ds}), has been proposed as a viable solution of the dark matter/dark energy puzzle. According to this theory, dark matter and dark energy can be understood as an artefact, due to the attempt to describe global physical effect in purely local galactic terms.

Here we study BH thermodynamics in CG. We do so by employing various methods, as, e.g., \textit{Thermodynamic Geometry} (TG) \cite{Rao,Ruppeiner:1979,Ruppeiner:1991,Ruppeiner:1998,Ruppeiner:1995zz}, that is also a very active area of research by itself. Below we show that such approach can fruitfully be applied also in CG, an interesting result in its own right.

Our focus here is on stability and phase transitions. Hawking first suggested \cite{Hawking:1974and1975and1976} that the Schwarzschild BH is thermodynamically unstable, due to its \textit{negative} specific heat, for which the more the BH emits, the hotter becomes. Subsequent work shows that other BHs in 3+1 dimensions, such as Kerr, Reissner-Nordstr\"{o}m and Kerr-Newmann, are unstable, i.e. the entropy is a concave function. As we shall see, it is not clear whether this condition is sufficient to characterize the stability of non-extensive systems, like BHs.

Currently, two are the main approaches to have BHs with positive specific heat. The first is to lower dimensions to 2+1, hence to consider the Ba\~{n}ados-Teitelboim-Zanelli (BTZ) solution there \cite{BTZ1992}. The other way is to include a negative cosmological constant, $\Lambda < 0$, that is to have asymptotically Anti de Sitter (AdS) spacetimes rather than flat Minkowski~\cite{Hawking:1982dh}.

Relation between stability and positive values of the specific heat holds only for extensive systems~\cite{CallenBOOK}. However, BH are not extensive objects, their entropy scales as the area, not the volume, hence stability is not related to the concavity of the entropy function. Stability of this kind of systems need be studied with a different criteria, named \textit{Poincar\'{e}} (or \textit{turning point}) method~\cite{Poincare}, that holds for a generic out-of-equilibrium entropy, and requires only that equilibrium is reached in an extreme point~\cite{Friedman:1988er,Katz:1993up,Okamoto:1994aq,Parentani:1994wr,Kaburaki:1993ah,Parentani:1994aa,Arcioni:2004ww}. The peculiarity of gravitating systems, in this respect, is that the algebraic sign ($\pm$) of the thermodynamic quantities is not an indicator of stability. On the other hand, a \textit{change of sign} is correlated to a \textit{change of stability}. This analysis is not always compatible with the conclusions based on the concavity of the entropy. For example, the Poincar\'{e} method does not predict instability for the Schwarzschild or Kerr BHs~\cite{Arcioni:2004ww}.

In addition to stability, phase transitions in BHs are of great interest. Important examples are the Hawking-Page phase transition~\cite{Hawking:1982dh}, the AdS/CFT correspondence and its relation to deconfinement at finite temperature in gauge theories. A simple way to study BH phase transitions is to employ the TG proposed by Ruppeiner. TG is based on the definition of a metric in the thermodynamic space~\cite{Rao,Ruppeiner:1979,Ruppeiner:1991,Ruppeiner:1998,Ruppeiner:1995zz}
\begin{equation*}
g^{\rm TG}_{\mu\nu} \equiv -\frac{\partial^2 S}{\partial X^\mu \partial X^\nu}
\;.
\end{equation*}
where $S$ is the entropy, while $X^\mu\in(E,P,N,\,\cdots)$, with $E$ energy, $P$ pressure, $N$ number of particles, etc.. This way, critical phenomena are related to distinctive signs of the scalar  curvature, $R^{\rm TG}$, obtained from such metric: $R^{\rm TG} = 0$ means a system made of noninteracting components, while for $R^{\rm TG} < 0$ such components attract each other, and for $R^{\rm TG} > 0$ repel each other. Moreover, $R^{\rm TG}$ diverges in a second order phase transition as the correlation volume, while it appears to have a local maximum at a crossover, as happens in quantum chromodynamics~\cite{Castorina:2018gsx,Castorina:2018ayy,Castorina:2019jzw,Zhang:2019neb}.

TG has been tested in many different systems: phase coexistence for helium, hydrogen, neon and argon~\cite{Ruppeiner:2011gm}, for the Lennard-Jones fluids~\cite{May:2012,May:2013}, for ferromagnetic systems and liquid-liquid phase transitions~\cite{Dey:2011cs}; in the liquid–gas-like first-order phase transition in dyonic charged AdS BH~\cite{Chaturvedi:2014vpa}; in quantum chromodynamics (QCD) to describe cross-over from Hadron gas and Quark-Gluon-Plasma~\cite{Castorina:2018gsx,Castorina:2018ayy,Castorina:2019jzw,Zhang:2019neb}; in the Hawking-Page transitions in Gauss–Bonnet–AdS~\cite{Sahay:2017hlq}, Reissner-Nordstrom-AdS and the Kerr-AdS ~\cite{Sahay:2010tx}. A list of results have been obtained by applying TG to BHs~\cite{Aman:2003ug,Shen:2005nu,Aman:2005xk,Ruppeiner:2007hr,Ruppeiner:2008kd,Sarkar:2008ji,Bellucci:2011gz,Wei:2012ui,Wei:2015iwa,Sahay:2016kex,Ruppeiner:2018pgn}.

Our paper is organized as follows. We shall first briefly review the basic ideas of CG and of BHs in CG in Section~\ref{sec:BHC}. We shall then introduce Ruppeiner’s TG in Section.~\ref{sec:Rup}, while Section~\ref{sec:STABILITY} offers a discussion about stability in extensive and non-extensive systems. Section~\ref{sec:StabCBH} is devoted to the main results of this paper, that is the study of the stability and phase transitions of the (A)dS Schwarzschild BH in CG. Our comments and conclusions are in Section~\ref{sec:CC}. The paper is closed by two Appendices, devoted to the details of some computations.

\section{\label{sec:BHC}BH Thermodynamics in CG}

Let us start by recalling what is Weyl symmetry, as this will help us to clarify the general set-up, and to identify the theory of gravity we are actually investigating. A scaling of the distance
\begin{equation}\label{norm}
  \|x\|^2 \equiv g_{\mu \nu} x^\mu x^\nu \, \to \, \Omega^2 \|x\|^2
\end{equation}
can be obtained in two ways
\begin{equation}\label{Xscaling1}
x^\mu \to \Omega \; x^\mu \quad {\rm and} \quad g_{\mu \nu} \to g_{\mu \nu} \,
\end{equation}
or
\begin{equation}\label{Xscaling2}
x^\mu \to x^\mu \quad {\rm and} \quad g_{\mu \nu} \to \Omega^2 g_{\mu \nu} \,.
\end{equation}
In both cases the matter fields $\Phi_i$, when present, transform as
\begin{equation}\label{Xscaling3}
\Phi_i (x) \to \Omega^{d_\Phi} \Phi_i (x) \,,
\end{equation}
where $d_\Phi$ is their scale dimension.

The two scalings are equivalent in practise, but not in spirit. The second way to transform are the \textit{Weyl transformations}. They turn a spatiotemporal transformation (acting on $x^\mu$) into a field/internal transformation (acting on $g_{\mu \nu}$), hence something that, when $\Omega (x)$, one could gauge in the usual way, i.e., by improving standard derivatives (e.g., $\partial_\mu \phi$, for $\Phi_i \equiv \phi$, a scalar field) to covariant derivatives (e.g., $\partial_\mu \to \partial_\mu + d_\phi W_\mu$) with the Weyl-gauge field transforming as $W_\mu \to W_\mu - \partial_\mu \ln \Omega$. On this see \cite{lor}, and also \cite{LorWeylHistory}.

The latter reference tells the history of this idea, that actually is the first historical example of what we nowadays call \textit{gauge theory}. As recalled there, Weyl's first attempt to propose a theory of gravity based on this symmetry was not taken seriously by Einstein (nor other scientists of that time). Over the years, though, the role of Weyl symmetry in gravity had a resurgence, starting from the work of Deser, Jackiw and Templeton in 2+1 dimensions in \cite{DJT}, implicitly as part of a larger theory, then more explicitly as a separate subject, again in 2+1 dimensions in \cite{hornewitten} (where the term ``conformal gravity'' is first used), and later in 1+1 dimensions in \cite{ioriojackiw}. On the other hand, a \textit{dilaton-improved} Einstein-Hilbert action in 3+1 dimensions, that enjoys Weyl symmetry, is employed in \cite{leslaw1e2} to study the thermodynamics of BHs. There too such theory is called \textit{conformal gravity}.

For us, CG (which we may as well call \textit{Weyl gravity}) is a theory of gravity in 3+1 dimensions, enjoying Weyl symmetry, whose action is~\cite{Riegert:1984zz,Mannheim:1988dj,Klemm:1998kf,Mannheim:2005bfa,Mannheim:2011ds}
\begin{equation}\label{Iw}
I_{W} =  \alpha_{W} \;\int d^4x \sqrt{|g|}\;C^{\mu\nu\rho\sigma} C_{\mu\nu\rho\sigma}
\;,
\end{equation}
where
\begin{equation}
\begin{split}
& C^\lambda_{\mu\nu\rho}
=
R^\lambda_{\mu\nu\rho}
+
\frac{1}{6}\;R \left(g^\lambda_{\nu}R_{\mu\rho}-g^\lambda_{\rho}R_{\mu\nu}\right)-\\
&
-
\frac{1}{2}\left(
g^\lambda_{\nu}R_{\mu\rho}
-
g^\lambda_{\rho}R_{\mu\nu}
-
g_{\mu\nu}R^\lambda_{\rho}
+
g_{\mu\rho}R^\lambda_{\nu}
\right)
\end{split}
\end{equation}
is the  Weyl tensor, expressed in terms of the Riemann tensor, $R^\lambda_{\mu \nu \rho}$, and its contractions, the Ricci tensor, $R_{\mu \nu}$ and the scalar curvature, $R$.

In dimensions $n>3$, under Weyl transformations: $C^\lambda_{\mu\nu\rho} \to C^\lambda_{\mu\nu\rho}$ (that, from $C^\lambda_{\mu\nu\rho} (\eta) = 0$, gives $C^\lambda_{\mu\nu\rho} (\Omega^2 \eta) = 0$), which, together with $g_{\mu \nu} \to \Omega^2 g_{\mu \nu}$, and $g^{\mu \nu} \to \Omega^{-2} g^{\mu \nu}$, gives $C^{\lambda \mu\nu\rho} \to \Omega^{-6} C^{\lambda \mu \nu \rho}$ and $C_{\lambda \mu\nu\rho} \to \Omega^{2} C_{\lambda \mu \nu \rho}$, while $\sqrt{|g|} \to \Omega^n \sqrt{|g|}$. With these
\begin{equation}\label{Iwscaling}
  I_W \to \Omega^{d_\alpha + n - 4} I_W \,,
\end{equation}
where $\alpha_W \to \Omega^{d_\alpha} \alpha_W$  (recall that $d^n x$ does not scale \footnote{Notice, though, that $x_\mu = g_{\mu \nu} x^\nu \to \Omega^2 g_{\mu \nu} x^\nu = \Omega^2 x_\mu$.}). This means that $d_\alpha = 4 - n$, i.e., for $n=4$, our case, $\alpha_W$ is dimensionless. From a different perspective, since $[C^\lambda_{\mu\nu\rho}] = [L]^{-2}$ (recall that $[g_{\mu \nu}] = 1$), then, in units $c = 1 = \hbar$, $[\alpha_W] = [L]^{(4 - n)}$, that is again the same result.

Variation of the action (\ref{Iw}) with respect to $g_{\mu\nu}$ gives the Euler-Lagrange equations for the theory, called \textit{Bach equations} in vacuum (no matter)
\begin{equation}
\frac{g_{\mu\alpha}g_{\nu\beta}}{\sqrt{|g|}}\;\frac{\delta I_W}{\delta g_{\alpha\beta}} = 2\;\alpha_{W}\;W_{\mu\nu} = 0
\;,
\end{equation}
where the \textit{Bach tensor} is defined as
\begin{equation}
W_{\mu\nu}
=
W^{(2)}_{\mu\nu} - \frac{1}{3}\;W^{(1)}_{\mu\nu}
\;,
\end{equation}
with
\begin{equation}
W_{\mu\nu}^{(1)}
=
2\;g_{\mu\nu} R^{;\beta}_{\;;\beta} -
2\; R_{;\mu\;;\nu}
-2\; R R_{\mu\nu} + \frac{1}{2} \; g_{\mu\nu} R^2
\;,
\end{equation}
\begin{equation}
\begin{split}
W_{\mu\nu}^{(2)}
=&
\frac{1}{2}\;g_{\mu\nu} R^{;\beta}_{\;;\beta}
 +
R_{\mu\nu\;;\beta}^{\;\;\;;\beta}
-
R_{\mu\;;\nu;\beta}^{\;\;\beta}
-
R_{\nu\;;\mu;\beta}^{\;\;\beta}
\\
&-
2\;R_{\mu\beta}R_{\nu}^{\;\beta} +\frac{1}{2}\;g_{\mu\nu} R_{\alpha\beta}R^{\alpha\beta}
\;,
\end{split}
\end{equation}
where, as usual, $\bullet ; \mu \equiv \nabla_{\mu} \bullet$, indicates the covariant derivative.

The complete and exact vacuum solution is a static, spherically symmetric geometry given by~\cite{Mannheim:1988dj}
\begin{equation}\label{eq:ds2}
ds^2 = -f(r)\;dt^2 + \frac{dr^2}{f(r)} + r^2\;d\omega^2_{S^2}
\;,
\end{equation}
where $r$ is the radial coordinate, and $d\omega^2_{S^2}$ is the line element of the $2$-sphere $S^2$. All other static, spherically symmetric line elements are conformal to (\ref{eq:ds2}). The function $f(r)$ takes the form~\cite{Riegert:1984zz,Mannheim:1988dj,Klemm:1998kf}
\begin{equation}\label{eq:EH}
f(r) = c_1 + c_2 \; r + \frac{c_3}{r} +c_4\;r^2
\;,
\end{equation}
where $c_1$, $c_2$, $c_3$ and $c_4$ are four integration constants, of which only three are independent, due to the constraint
\begin{equation}\label{eq:constrain}
c_1^2 = 3\;c_2\;c_3+1 \;,
\end{equation}
imposed by Bach equations. Indeed, since all information is contained in $W^{rr}$, which only involves derivatives up to third order. In a static geometry $W^{0r}$ is identically zero, while other terms are obtained through the Bianchi and trace identities~\cite{Mannheim:1988dj}.

The event horizon is found in the usual way, by solving $f(r_h) = 0$. We do not deem useful to show the complicated general solution here, but we shall discuss the particular cases of interest, when they appear.

\subsection{Comparison with standard gravity}

We need now to understand which aspects of interests change, and which do not change in this picture, as compared to standard gravity of the Einstein kind (including a cosmological constant). Our concerns are (a) the integration constants $c_1, ..., c_4$, that need be associated to BH thermodynamic variables, such as mass, volume, pressure, etc.; (b) the BH temperature; (c) the BH entropy, that plays a crucial role in TG.

(a) First, notice that four scales have appeared in (\ref{eq:EH}), and we are soon going to identify them with scales known from Einstein gravity, such as, e.g., the cosmological constant. This is quite a different story, tough. In the non-conformal case, one starts off with an action that indeed has the scale. Hence, clearly, this scale is found in the solutions. Here such scales are \textit{dynamically generated}, i.e., they only appear when the Euler-Lagrange equations are solved. Since such equations identify the field configurations that minimize the action functional (Hamilton principle), this amount to fix a specific minimum, identified by a specific set of constants. On the other hand, conformal/Weyl symmetry is at work also on-shell. Therefore, the scale fixing is only apparent, and, in a way reminiscent of the spontaneous breaking of symmetry: all configurations obtained Weyl-transforming the solution in (\ref{eq:EH}) are solutions too. Notice that this is an entirely classical phenomenon.

Enthalpy, evaluated as the Noether charge corresponding to the time-like Killing vector for (\ref{eq:ds2}), is~\cite{Lu:2012xu}
\begin{equation}\label{eq:E}
H =
\alpha_W \left[\frac{c_2\;(c_1 - 1)}{6}
-
c_3\;c_4 \right]
\;.
\end{equation}
We observe that, if in \eqref{eq:EH} one sets
\begin{equation}\label{eq:c1} c_1 \equiv 1 \; \;, \; c_2 \equiv 0 \; \;, \; c_3 \equiv -\eta \; \;, \; \displaystyle c_4 \equiv -\frac{\Lambda}{3} \;,
\end{equation}
the line element (\ref{eq:ds2}) is an (A)dS line element, with~\cite{Hawking:1982dh},
\begin{equation}
	f(r) = 1 -\frac{\eta}{r} - \frac{\Lambda}{3}\;r^2\;,
\end{equation}
where $\eta$ in Einstein gravity is $2 M G$, while $\Lambda$ plays the role of the cosmological constant. Immediately, one sees that the dimensions are not the same here, as compared to Einstein gravity, essentially due to the lack of the Newton constant $G$ here, with $[G] = [L]^2$. In fact, while here $[\eta] = [L]$ matches $[M G] = [L]$ of Einstein gravity, the same cannot be said of $M$ itself, for which $[M] = [L]^{-1} \neq [\eta] = [L]$ (mismatch of $[L]^2$). Similarly, here $[\Lambda] = [L]^{-2}$, just like in Einstein gravity, nonetheless, when one computes the associated \textit{pressure}, say it $P$, according to Einstein gravity one finds $P \sim \Lambda/G$, hence $[P] = [L]^{-4}$ (again a mismatch of $[L]^2$). Nonetheless, if one considers that, as we shall see, the variable playing the role of the \textit{volume} here (we call it $\Gamma$ below) is such that $[\Gamma] = [L]$, rather than $[V] = [L]^3$ of Einstein gravity (mismatch of $[L]^2$), we see that a proper ``$PdV$'' can be included in our thermodynamic analysis as ``$(-\Lambda) d\Gamma$'' (on the sign, see below).

(b) As for the Hawking temperature, although its relation with the ``mass'' is clearly modified by the new settings, as explained above, its structure is insensitive to conformal/Weyl transformations. This is an old, and quite general result proved in \cite{jacobson}, based on the fact that a conformal/Weyl invariant surface gravity, $\kappa_W$, of a conformal Killing horizon can be found to agree with the surface gravity, $\kappa$, of the Killing horizon. The latter is proportional to the Hawking temperature in the usual way, $T = \kappa / 2 \pi = \kappa_W / 2 \pi$, where the last equality is the result. More recently, this was applied to certain (2+1)-dimensional systems, to show how an ultra-static metric can still have interesting conformal Killing horizons, hence, an associated Hawking phenomenon \cite{Beltrami1,Beltrami2}.

Let us then evaluate the Hawking temperature according to
\begin{equation}\label{eq:T0}
T = \frac{1}{4\pi}\; \left[ \frac{df(r)}{dr}\right]_{r = r_h} = \frac{r^2_h-\left(3\;c_3+c_1\;r_h\right)^2}{12\;\pi\;c_3\;r^2_h}
\;,
\end{equation}
and stress that this expression is invariant under Weyl transformations, $T \to T$.

(c) The last, and most important point to discuss is the BH entropy, that is central to the analysis we want to perform, based on TG. One could be tempted to use the Bekenstein formula for it
\begin{equation}\label{bekensteinS}
  S_B = \frac{1}{4 G} {\cal A}_\Sigma \;,
\end{equation}
as sometimes done in the literature, see, e.g., \cite{leslaw1e2}, for similar theories that are different from the one we deal with here, as explained earlier here. There ${\cal A}_\Sigma$ is the area of the event horizon $\Sigma$.

If this entropy is used, we have immediately two issues. First, we have no scale $G$ in our theory here. Second, the area must respond to Weyl transformations, hence, recalling that out of this entropy we need to construct a metric in the thermodynamic space of TG, such metric would scale too.

Both problems are solved at once if one considers that this is not Einstenin gravity, and a generalization of Bekenstein entropy to gravity theories other than Einstein's is indeed available. Such generalization was discovered by Wald in \cite{wald}, and it is based on the Noether charge associated to diffeomorphisms, that are generic translations. In this approach, the expression for the entropy is the integral of the Noether potential over the BH’s bifurcation surface \cite{wald}, or any arbitrary slice of a Killing horizon  \cite{jacobsonANDwald} (so that the formula applies to BHs other than the eternal stationary, including those formed through a dynamical process in the past). For a recent discussion see, e.g., \cite{SwaldLQG}.

The explicit formula is \cite{wald} \cite{SwaldLQG}
\begin{equation}\label{SwaldGeneral}
  S = - 2 \pi \int_\Sigma d^2 x \; \sqrt{h} \; \epsilon_{\mu \nu} \epsilon_{\rho \sigma} \; \displaystyle \frac{\delta {\cal L}}{\delta R_{\mu \nu \rho \sigma}} \;,
\end{equation}
where $\epsilon_{\mu\nu} = 1/\|x\|^2 (x_\mu x_\nu - x_\nu x_\mu)$ is the \textit{normal bivector} to the arbitrary slice of the Killing horizon $\Sigma$, identified by ($r=r_h$,  and  $t={\rm const.}$) (note that $\epsilon^{\mu\nu}\epsilon_{\mu\nu}=-2$), with induced metric $h_{\alpha \beta}$, whose determinant is $h$. The formula applies to theories with Lagrangian density ${\cal L} (R_{\mu \nu \rho \sigma}, \nabla_\lambda R_{\mu \nu \rho \sigma}, \nabla_{(\lambda \kappa \cdots )} R_{\mu \nu \rho \sigma}, \Phi_i, \nabla_\lambda \Phi_i, \nabla_{(\lambda \kappa \cdots)} \Phi_i)$, that, in general, includes the matter fields, $\Phi_i$, that also may contribute to the entropy.

When ${\cal L} = \displaystyle \frac{1}{16 \pi G}  R$, then $\delta {\cal L}/ \delta R_{\mu \nu \rho \sigma} = \displaystyle \frac{1}{32 \pi G} (g^{\mu \rho} g^{\nu \sigma} - g^{\nu \rho} g^{\mu \sigma})$, hence
\begin{eqnarray}\label{waldTObekenstein}
  S & = & - \displaystyle \frac{1}{16 G} \int_\Sigma d^2 x \; \sqrt{h} \; \epsilon_{\mu \nu} \epsilon_{\rho \sigma}  (g^{\mu \rho} g^{\nu \sigma} - g^{\nu \rho} g^{\mu \sigma}) \nonumber \\
  & = & \frac{1}{4 G} \int_\Sigma d^2 x \; \sqrt{h} = \frac{1}{4 G} {\cal A}_\Sigma = S_B \;.
\end{eqnarray}
That is, for Einstein theory the two entropies coincide (due to $\delta \Lambda / \delta R_{\mu \nu \rho \sigma} = 0$, this is true also in the case of a nonzero cosmological constant).

In our case, though, the two entropies differ, and the Wald entropy is clearly
\begin{equation}\label{eq:Sgeneral}
S = -4\,\pi \alpha_W \; \int_\Sigma \sqrt{h}\; d\Sigma C^{\mu\nu\lambda\rho} \; \epsilon_{\mu\nu}\epsilon_{\lambda\rho} \;.
\end{equation}
The first thing we notice is that this entropy is invariant under Weyl scaling, as it must be. Indeed, $C^{\mu\nu\lambda\rho} \to \Omega^{-6} C^{\mu\nu\lambda\rho}$, whereas $\sqrt{h} \to \Omega^2 \sqrt{h}$, and $\epsilon_{\mu \nu} \to \Omega^2 \epsilon_{\mu \nu}$ (recall that $x_\mu \to \Omega^2 x_\mu$ and $\| x \| \to \Omega \| x \|$). Altogether, $S \to \Omega^{2} \Omega^{2}\Omega^{2} \Omega^{-6} S = S$.

Writing in explicitly the Weyl tensor evaluated for the geometry (\ref{eq:ds2}) with (\ref{eq:EH}), one obtains~\cite{Cognola:2011nj,Lu:2012xu}
\begin{equation}\label{eq:S}
S = \frac{4 \pi \alpha_W }{6} \left(1 - c_1 - \frac{3\;c_3}{r_h} \right) \;.
\end{equation}
where $\Sigma \equiv S^2$, is the two sphere, so that we have the factor $4 \pi$.

A note on the coupling $\alpha_W$ is now in order. Putting the choice (\ref{eq:c1}) into \eqref{eq:S}, one finds that the entropy  is positive only if
\begin{equation}
\alpha_W\;c_3<0 \;.
\end{equation}
Since in Einstein gravity $\eta = 2 M G^2>0$, the Einstein-Schwarzschild limit ($c_1=1$ and $c_2=0$) also requires $c_3<0$. Therefore, to have a positive entropy in the Schwarzschild limit, the coupling constant $\alpha_W$ must be positive. Moreover, from eq.~\eqref{eq:E} the entalphy of a Schwarzschild BH is positive if $\alpha_W c_3\, c_4<0$, that is  $c_4>0$ (i.e $\Lambda<0$), and negative otherwise.

The enthalpy, the temperature, and the entropy in Eqs.~(\ref{eq:E}), (\ref{eq:T0}), and (\ref{eq:S}) are evaluated geometrically. However, they have a clear thermodynamic meaning. Moreover, in Weyl gravity the cosmological constant arises as a parameter of the solution rather than as a fixed parameter of the theory and thus it is natural to treat it as a potential.

\subsection{\label{sec:Th}Thermodynamics}

According to the previous discussions and since there are three independent integration constants in \eqref{eq:EH}, the first law of BH thermodynamics is given by
\begin{equation}
dH = T\;dS + \Psi\;d\Xi + \Gamma\;d(-\Lambda)
\;,
\end{equation}
where the thermodynamic potentials
\begin{equation}\label{eq:V1}
\Xi \equiv c_2 \;, \quad \Lambda = - 3\,c_4\;
\end{equation}
and the variables~\cite{Lu:2012xu}
\begin{equation}\label{eq:V2}
\displaystyle \Psi = \frac{\alpha_W}{6} \; (c_1-1)
\;, \quad \displaystyle \Gamma = - \frac{\alpha_W}{6} \;c_3  \;,
\end{equation}
have been defined.

Without loss of generality, we set $\alpha_W = 1/(2\,\pi)$ (for $\alpha_W<0$ one has to change the sign of the rescaling also, $S\rightarrow - 2 \pi \, \alpha_W\, S$, in the next equations in order to have a positive entropy, but in this case one can study dS BHs only, see Appendix~\ref{app:a2}).

By writing eqs.~\eqref{eq:T0} and~\eqref{eq:S} in terms of the thermodynamic variables, one finds that the entropy depends only on the temperature, $T$, and on $\Gamma$,
\begin{equation}\label{eq:S3}
(3\;S-1)^2
 =
 1+\left(12\pi\right)^2\;\Gamma\;T
\;,
\end{equation}
and the two solutions, $S_+$ and $S_-$, are
\begin{equation}\label{eq:S4}
S_{\pm} = \frac{1\pm \sqrt{1+\left(12\pi\right)^2\Gamma T}}{3}
=
-4\;\pi\;\Psi + \frac{12\;\pi\;\Gamma}{r_\pm}
\;,
\end{equation}
where the last equality is due to the Wald formula of  eq.~\eqref{eq:S}, with $1/r_\pm$ solutions of eq.~\eqref{eq:T0}, i.e.
\begin{equation}\label{eq:rpm}
\frac{1}{r_{\pm}}
=
\frac{1+12\;\pi\;\Psi \pm \sqrt{1+\left(12\pi\right)^2\Gamma\;T}}{36\;\pi\;\Gamma}
\;.
\end{equation}
Therefore, the event horizon, $r_h$, is given by
\begin{equation}
r_h
=
\begin{cases}
r_- \;,& \text{for} \; S_-\\\\
r_+ \;,& \text{for} \; S_+
\end{cases}
\;.
\end{equation}
Note that $S_- < 1/3$, while $S_+ > 1/3$ $\;\forall \,\Gamma\,T$.

\begin{figure}
	\centering
	\includegraphics[width=0.8\columnwidth]{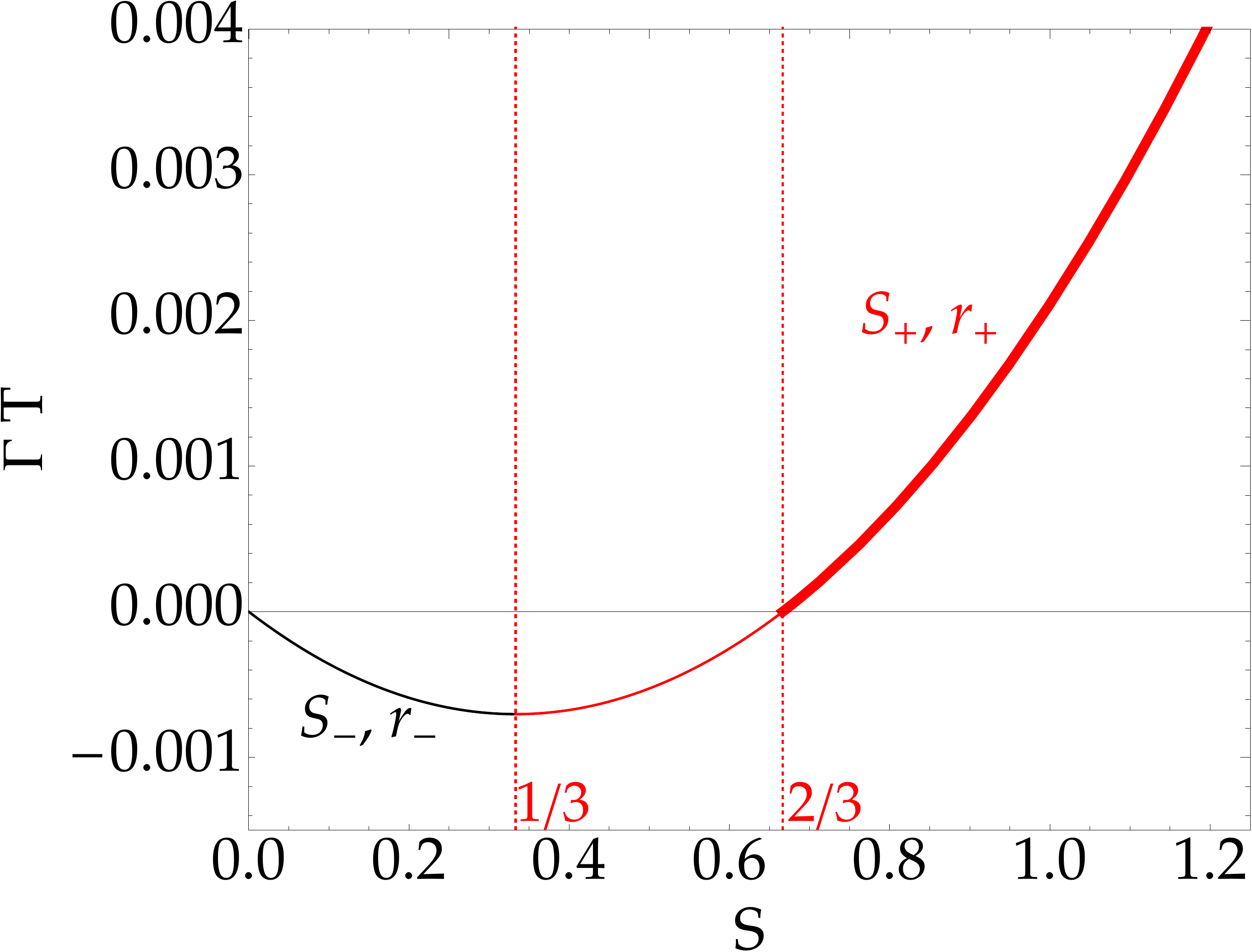}
	\caption{$S$ as a function of $\Gamma T$ from eq.~\eqref{eq:S3}: if the entropy is greater than $1/3$ the solution is given by $S_+$ and $r_+$, otherwise by $S_-$ and $r_-$. In bold we indicate the only physically acceptable values with $\Gamma T > 0$. }
	\label{fig:a1}
\end{figure}

It is useful to stress that the two branches, ($S_+,r_+$) and ($S_-,r_-$), should be considered completely independent. From now on, the sign + or - indicates the specific branch. The expressions of $\Lambda$, $H$ and $\Xi$ as a function of $T$, $\Gamma$ and $\Psi$ are in app~\ref{app:conti}.

However, the entropy as a function of $H$, $\Lambda$ and $\Xi$ (both for $(S_-,H_-,\Lambda_-,\Xi_-)$ or $(S_+,H_+,\Lambda_+,\Xi_+)$) is solution of the some
equation
\begin{equation}\label{eq:S2}
\begin{split}
&\Lambda_\pm \left[(1-S_\pm)\;S_\pm^2+32\;\pi^3\;\Psi(H_\pm, \Lambda_\pm,\Xi_\pm)^3\right]^2
=\\
&\quad
=-
12\;\pi^2\;H_\pm^2\;\Bigg[(S_\pm-1)\;S_\pm^2 +\\
&
\quad
+ 16\;\pi^2\;\Psi(H_\pm, \Lambda_\pm,\Xi_\pm)^2\;(1+4\;\pi\;\Psi(H_\pm, \Lambda_\pm,\Xi_\pm))\Bigg]
\;,
\end{split}
\end{equation}
where $\Psi(H_\pm, \Lambda_\pm,\Xi_\pm) $  can be evaluated by reversing
the Smarr relation~\cite{Lu:2012xu},
\begin{equation}\label{eq:Smarr}
H_\pm = \Psi\;\Xi_\pm - 2\;\Lambda_\pm\;\Gamma
\;,
\end{equation}
together with
\begin{equation}
\Xi_\pm = -\frac{2\;\Psi\;(1+6\;\pi\;\Psi)}{3\;\Gamma}
\;.
\end{equation}
From now on, the subscript ``$\pm$'' is not shown when the expression is valid for both sets $(S_-, H_-, \Lambda_-, \Xi_-)$ and  $(S_+, H_+, \Lambda_+, \Xi_+)$. Moreover, when
$S < 1/3$, what follows refers to the branch $S=S_-$, whereas when $S>1/3$, the results are for $S=S_+$.

We study the Schwarzschild (A)dS BH limit, i.e. the solution with $(c_1,c_2)=(1,0)$, or equivalently,  with  $(\Psi,\Xi)=(0,0)$, for which the equation~\eqref{eq:S2} becomes
\begin{equation}\label{EoS1}
\Lambda\;S^2\;(S-1)+12\;\pi^2\;H^2=0
\;.
\end{equation}
According to eq.~\eqref{EoS1},  $\Lambda <0$ when $S>1$ and $\Lambda >0$ if $0<S<1$. Therefore, AdS BHs have entropy larger than $1$ and positive $H$, the dS BHs have entropy $0<S<1$, and $H<0$.

On physical grounds, one can only consider the solution $S_+$ with $\Gamma T$ positive. Indeed, looking at the Schwarzschild limit, $\Psi=0$, of eq.~\eqref{eq:rpm} one finds that $\Gamma$ must be always positive in order to have a positive event horizon $r_\pm$. Thus for entropy $0<S<2/3$ the temperature is negative for positive ``volume'' $\Gamma$. The solutions are plotted in fig.~\ref{fig:a1}.

Let us summarize the previous results for the (A)dS Schwarzschild BHs as follows:
\begin{itemize}
		\item BHs with entropy $0<S<2/3$ have negative temperature $T$, and we do not study them, as we see no physical reason for $T<0$ in a classical scenario;
        \item excluding the branch $T<0$ means, at once, that a) at $T=0$ the BH has a relic entropy, $S=2/3$, and b) no value of the allowed $T$ corresponds to several values of $S$ (see \cite{Caldarelli_1999});
		\item BHs with entropy $2/3<S<1$ have positive $\Gamma$, $T$ and $\Lambda$ (they are dS BHs), but have negative enthalpy, $H<0$, because $sg(H)=-sg(\alpha_W\,\Gamma\,\Lambda)$;
		\item BHs with entropy $S>1$ have positive $\Gamma$, $T$ and $H$, but have $\Lambda<0$ (they are AdS BHs).
\end{itemize}

\section{\label{sec:Rup}Ruppeiner metric}

In the Gaussian thermodynamic fluctuation theory (see sec.~\ref{sec:STABILITY}), the probability of fluctuations away from the equilibrium configurations of a given system turns out to be~\cite{Ruppeiner:1979,Ruppeiner:1991,Ruppeiner:1998,Ruppeiner:1995zz}
\begin{equation}\label{eq:PGT}
dP\propto
\exp\left\{-\frac{\Delta \ell^2}{2}\right\} \;\;dX^1\cdots dX^n
\;,
\end{equation}
where $X^\mu$ are the usual variables describing the state of the system, i.e.  $X^\mu\in(E,P,N,\,\cdots)$, where $E$ is the energy, $P$ is the pressure, $N$ is the number of particles, etc. and
\begin{equation}
\Delta\ell^2 = g^{\rm TG}_{\mu\nu}\;\Delta X^\mu \;\Delta X^\nu
\;,
\end{equation}
with
\begin{equation}\label{eq:g1}
g^{\rm TG}_{\mu\nu} \equiv -\frac{\partial^2 S}{\partial X^\mu \partial X^\nu}
\;.
\end{equation}
Here $\Delta X^\mu = X^\mu - X^\mu_0$, represent fluctuations of the state variables  from an equilibrium configuration, $X^\mu_0$.

Consequently, the thermodynamic space can be equipped with a metric, $g^{\rm TG}_{\mu\nu}$, with a clear physical meaning: it measures  the distance between two states of the system and, through eq.~\eqref{eq:PGT}, it quantifies the probability of fluctuation between them. The farther the states are, the less likely they are to fluctuate.

In Ruppeiner theory  $\Delta \ell^2$ is  an invariant quantity and  can be evaluated by different coordinates. Indeed, the inverse metric, $g_{\rm TG}^{\mu\nu}$, can be written as~\cite{Sahay:2016kex}
\begin{equation}\label{eq:g2}
g_{\rm TG}^{\mu\nu} = \frac{\partial^2 \Phi}{\partial \theta_\mu \partial \theta_\nu}
\;,
\end{equation}
where the Massieu function $\Phi$  is the Legendre transform of the entropy with respect to the variables $\theta_\mu$:
\begin{equation}\label{eq:Massieu}
\Phi = S - \theta\mu X^\mu
\;.
\end{equation}
$X^\mu$ and $\theta_\mu$ are conjugate variables defined as
\begin{equation}\label{eq:cv}
\theta_\mu = \frac{\partial S}{\partial X^\mu}
\;,\qquad
X^\mu = -\frac{\partial \Phi}{\partial \theta_\mu}
\;.
\end{equation}
For example, if $X^\mu=(E,\,P,\,N,\cdots)$, then $\theta_\mu = \left(1/T,-V/T,-\mu/T,\cdots\right)$, where $V$ is the volume, $\mu$ the chemical potential, etc.

Once we have a metric, the next natural step is to evaluate the Riemann tensor and its contractions, and to ask for their physical relevance.

For systems depending only on two variables, $(X^1,X^2)$, the scalar curvature takes the simple form
\begin{equation}\label{eq:R}
R^{\rm TG} =
\frac{1}{2\;g_{\rm TG}^2}\;\left|
\begin{array}{ccc}
S_{,11} & S_{,12} & S_{,22}\\
S_{,111} & S_{,112} & S_{,122}\\
S_{,112} & S_{,122} & S_{,222}\\
\end{array}
\right|\;,
\end{equation}
where $g^{\rm TG}$ is the determinant of the metric, and the commas indicate standard derivatives, as usual.

One of the most important, widely checked, results of Ruppeiner geometry concerns the sign of the scalar curvature which is related   with the nature of the interaction occurring in the system.

According to the conventions of Eq.~\eqref{eq:g1} (notice that one could define the metric as $g^{\rm TG}_{\mu\nu} \equiv +\partial^2S/\partial X^\mu\partial X^\nu$, thus reversing the sign of $R^{\rm TG}$), the properties of $R^{\rm TG}$ are summarized as
\begin{itemize}
	\item $R^{\rm TG} = 0$, no interaction;
	\item $R^{\rm TG} > 0$, repulsive interaction dominates;
    \item $R^{\rm TG} < 0$, attractive interaction dominates.	
\end{itemize}
The magnitude of $R^{\rm TG}$ also bears useful information. Since this curvature scalar here has units of volume, Ruppeiner suggested that, at the critical points, it diverges as the correlation volume, $\xi^d$. This is the so-called ``interaction hypothesis''. In general, one can show that $R^{\rm TG}$ is a measure of the smallest volume where one can describe the given subsystem, surrounded by a uniform environment~\cite{Ruppeiner:1995zz}.

Therefore, the scalar curvature is a very useful tool to study phase transitions, especially when there is no knowledge of the underlying degrees of freedom, as is the case of BHs.

\section{\label{sec:STABILITY}Stability}

\subsection{Extensive thermodynamics}
In extensive thermodynamics, stability is defined through the Hessian of the entropy. Indeed, by considering for simplicity two systems in thermal contact with entropy $S(M,J)$ (where $M$ is the mass and $J$ is some other extensive parameter),  a transfer of some mass $dM$ from the first to the second subsystem,  would produce a new configuration with  total entropy given by \cite{CallenBOOK,Arcioni:2004ww}:
\begin{equation}\label{eq:St1}
S(M+dM,J)+S(M-dM,J)\leq 2\;S(M,J)
\;.
\end{equation}
The differential form of eq.~\eqref{eq:St1} is
\begin{equation}\label{eq:St2}
H_{,MM} = \frac{\partial^2 S}{\partial M^2} \leq 0
\;.
\end{equation}
Similarly, if one transfers  $dJ$, at fixed mass, from one subsystem to the other, one gets
\begin{equation}\label{eq:St3}
H_{,JJ} = \frac{\partial^2 S}{\partial J^2} \leq 0
\;,
\end{equation}
and in the case of  transfer of $dM$ and $dJ$
\begin{equation}\label{eq:St4}
\det H =  \frac{\partial^2 S}{\partial J^2}\;\frac{\partial^2 S}{\partial M^2}
-
\left(\frac{\partial^2 S}{\partial M\;\partial J}\right)^2
\geq 0
\;.
\end{equation}
Only two of eqs.~(\ref{eq:St2}-\ref{eq:St4}) are independent and the entropy extensivity  is a crucial hypothesis to obtain eq.~\eqref{eq:St1}. Thus, in ordinary thermodynamics the stability can be achieved by requiring two of eqs.~(\ref{eq:St2}-\ref{eq:St4}). This is not the case for non-extensive systems, such as BHs. However, before discussing how to infer stability indications in non-extensive systems, it is useful to recall the connection between stability and Ruppeiner metric.

Ruppeiner metric, defined in eq.~\eqref{eq:g1}, is essentially the opposite of the Hessian of the entropy and  a generalization of the condition~(\ref{eq:St2}-\ref{eq:St4}) for systems with more extensive variables can be done as follows.
Let us consider a system  with entropy $S$, in thermal contact and in equilibrium  with an environment, of entropy $S_e$. The total entropy is
\begin{equation}\label{eq:St5}
S_{tot} = S + S_e
\;.
\end{equation}
Due to  a small fluctuation $\left(\Delta X^\alpha,\;\Delta X_e^\alpha\right)$ away from the equilibrium values and since for isolated systems energy conservation requires $\Delta X=-\Delta X_e$,  the maximum entropy condition is equivalent to
\begin{equation}\label{eq:St6}
\frac{\partial S}{\partial X^\mu}
=
\frac{\partial S_e}{\partial X_e^\mu}
\;,
\end{equation}
and for large environments, the total entropy changes according to~\cite{Ruppeiner:2007hr}
\begin{equation}\label{eq:St7}
\begin{split}
\Delta S_{tot}
\simeq &
\frac{1}{2}\;\left(\frac{\partial^2 S}{\partial X^\mu\;\partial X^\nu}
\right)\;\Delta X^\mu \Delta X^\nu
=\\
=&
-
\frac{1}{2}\;g^{\rm TG}_{\mu\nu}\; \Delta X^\mu \Delta X^\nu
\;.
\end{split}
\end{equation}
Therefore, the Gaussian probability to find the state at $X^\mu+\Delta X^\mu$, starting from $X^\mu$, is
\begin{equation}\label{eq:St8}
\begin{split}
dP\propto\;&
\exp\left\{-\frac{g^{\rm TG}_{\mu\nu}}{2}\;\Delta X^\mu\;\Delta X^\nu\right\} \;\;dX^1\cdots dX^\mu
=\\
=&
\exp\left\{-\frac{\Delta \ell^2}{2}\right\} \;\;dX^1\cdots dX^\mu
\;,
\end{split}
\end{equation}
where $\Delta \ell^2 = g^{\rm TG}_{\mu\nu}\;\Delta X^\mu\;\Delta X^\nu$ is the distance between $X^\mu$ and $X^\mu+\Delta X^\mu$ in the thermodynamic manifold.

Eq.~\eqref{eq:St7} allows to easily generalize the conditions of eqs.~(\ref{eq:St1}-\ref{eq:St3}) to systems with more variables. After the fluctuations $\Delta X^\mu$ have taken place, the new state is less stable than the previous one, if and only if its entropy is smaller than, or equal to, the entropy of the initial one, so that $\Delta S_{tot}\leq0$. This implies that $g^{\rm TG}_{\mu\nu}$ must be a positive definite matrix, that can be studied, e.g., through the Sylvester criteria.

\subsection{Stability in non extensive thermodynamics}

Without requiring extensiveness of the potentials, the system stability can be studied by the Poincar\'{e} method~\cite{Poincare,Arcioni:2004ww} which is constructed in analogy with ordinary thermodynamics.

Let us consider a system out of equilibrium with entropy $\widehat S(Y^\mu,X^\mu)$, being $X^\mu=\{E,\;N\;,\cdots\}$ the usual equilibrium  variables (with  conjugate variables, $\theta_\mu$, as discussed in the previous section) and $Y^\mu$ other variables characterizing the non equilibrium condition. Clearly, equilibrium takes place when the system depends only on $X^\mu$, i.e., when~\cite{Katz:1993up,Okamoto:1994aq,Arcioni:2004ww}
\begin{equation}
Y^\mu = Y^\mu (X^\nu)
\;.
\end{equation}
These can be regarded as solution of the equilibrium equations
\begin{equation}\label{eq:St10}
\frac{\partial \widehat S}{\partial Y^\mu}\Bigg|_{eq}=0
\;.
\end{equation}
Let us define now some non-equilibrium functions (in the same way as in eq.~\eqref{eq:cv})
\begin{equation}\label{eq:St11}
\widehat \theta_\mu(Y^\nu,X^\nu) \equiv \frac{\partial \widehat{S}}{\partial X^\mu}
\;,
\end{equation}
and such that at the equilibrium
\begin{equation}
\widehat \theta_\mu(Y^\nu,X^\nu)\Bigg|_{eq}=\theta_\mu(X^\nu)
\;.
\end{equation}
If one assumes that $Y^\mu$ can be chosen  in
such a way the Hessian of $\widehat S$ is diagonal, the ``Poincarè coefficients of stability'' are defined through the equations
\begin{equation}
\lambda_\rho (X^\mu) \equiv \partial^2_{\rho}\widehat S(Y^\nu(X^\mu),X^\mu)\Bigg|_{eq}
\;.
\end{equation}
where $\lambda_\rho (X^\mu)$ are the Hessian eigenvalues and an instability appears if some $\lambda_\rho> 0$.

\begin{figure}
	\centering
	\includegraphics[width=0.8\columnwidth]{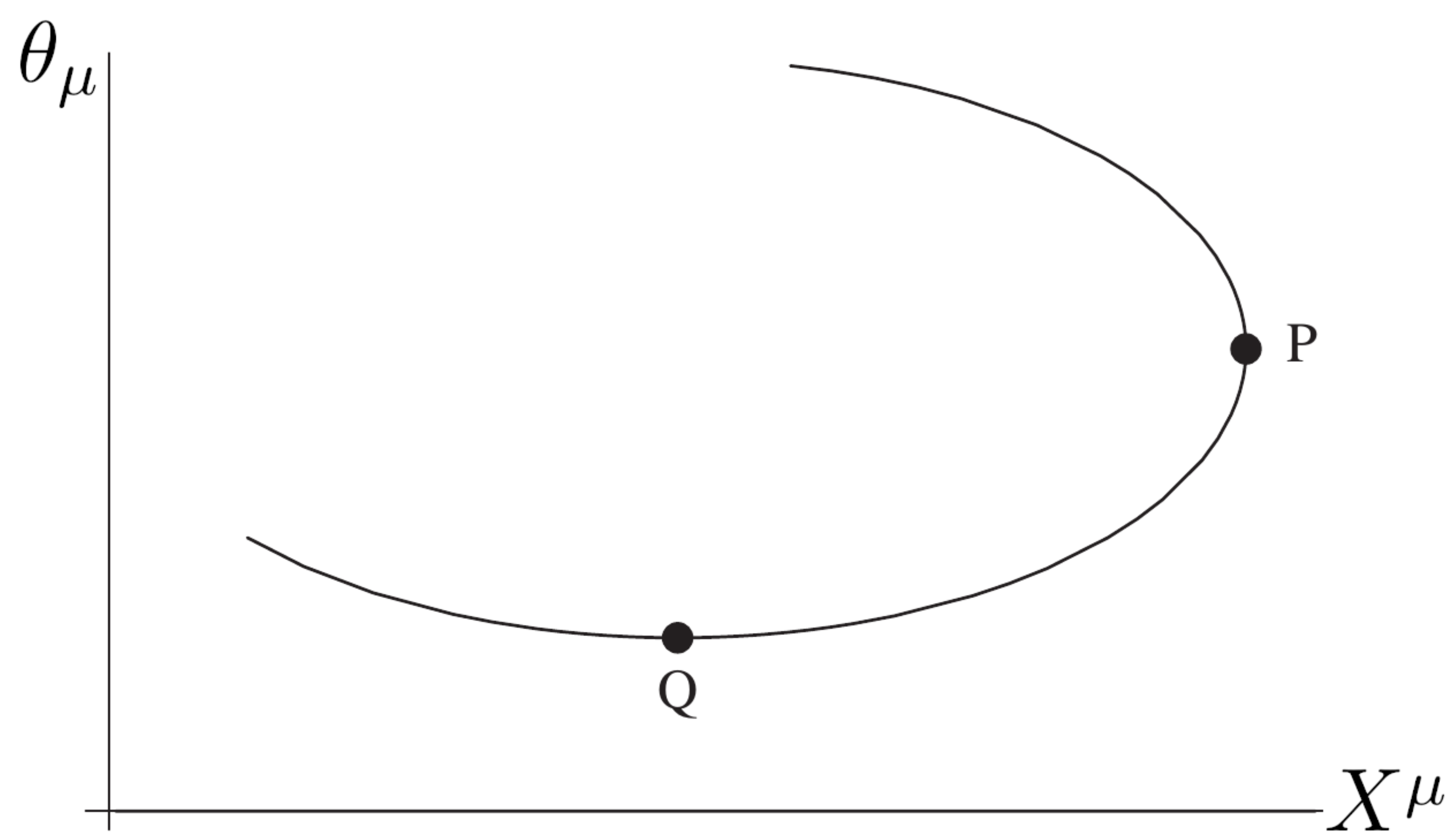}
	\caption{Generic plot of a conjugate variable $\theta_\mu$ against $X^\mu$, along an equilibrium sequence. The point P is a turning point. The upper branch is unstable,
		while the lower branch can be either fully stable, or simply more stable than the upper branch. The sign of the slope also changes at the horizontal tangent at $Q$\'{}, but this has no relation with a change of stability, even if the slope changes sign there. Figure from.~\cite{Arcioni:2004ww}.}
	\label{fig:St1}
\end{figure}

Without any knowledge about the out of equilibrium function, $\widehat S$, due to eqs.~(\ref{eq:St10},\ref{eq:St11}), one can link the equilibrium variables $\theta_\mu$ and $X^\mu$, with the non equilibrium ones by~\cite{Arcioni:2004ww}
\begin{equation}\label{eq:St12}
\frac{\partial \theta_\mu}{\partial X^\mu}
=
\left(
\frac{\partial^2 \widehat S}{\partial {Y^\mu}^2}
\right)_{eq}
-
\sum_\rho \;\frac{1}{\lambda_\rho}\; \left(
\frac{\partial \widehat \theta_\mu}{\partial Y^{\rho}}
\right)_{eq}^2
\;.
\end{equation}
The left-hand side involves only equilibrium variables, while the right-and side contains only the non equilibrium ones. Therefore, by previous equation, one obtains informations on the non equilibrium states if the properties of the system at equilibrium are known. For example, if the function $\theta_\mu = \theta_\mu(X^\mu)$ has an inflection point (point $P$ in fig.~\ref{fig:St1}) and $\partial \theta_\mu/\partial X^\mu$ changes sign, also the right-hand side of eq.~\eqref{eq:St12} changes its sign. This could be due to an eigenvalue turning from negative to positive value (or vice versa), describing a new phase in the stability of the system.

In fact, when at least one of the eigenvalues $\lambda_\rho$ changes sign, the Hessian has a zero, and $\partial \theta_\mu/\partial X^\mu$ diverges. This implies that the plot of $\theta^\mu(X^\nu)$ along the equilibrium points has a vertical tangent and one can study a change in stability by inspection of the plot of the equilibrium functions $\theta^\mu(X^\nu)$.

Moreover, one has to verify which branch is stable and in ref. ~\cite{Kaburaki:1993ah,Parentani:1994aa,Arcioni:2004ww} the following criteria have been suggested:

1) if one can prove the stability of even a single point, then all the other ones in the same stability sequence  are stable, until the first turning point is reached. After the turning point the system is unstable;

2) if a stable point is unknown, one can never say anything about the branch with positive slope. Instead, the branch with negative slope, near the turning point, is always unstable;

3) changes of stability can only occur at turning points or bifurcations. Indeed, according to eq.~\eqref{eq:St12}, there are other stability points besides the turning points, since it is possible that $(\partial_\rho \widehat \theta_\nu)_{eq}=0$ when the sign of $\lambda_\rho$ changes, but  $\partial_\mu \theta_\nu$ does not diverge. It can be shown that this can only happen at a bifurcation point ~\cite{Kaburaki:1993ah,Parentani:1994aa,Arcioni:2004ww};

4) a vertical asymptote signals the endpoints (boundary) of the (non)equilibrium sequence and it is not related to stability.

Finally, there are points where the slope of $\theta_\mu$ changes, but $\partial \theta_\mu /\partial X^\mu =0$ (see points $Q$ in figure~\ref{fig:St1}), but they do not correspond to system instability.
These points could indicate a sign variation in the specific heat, as in the case of the four dimensional Kerr BH in the microcanonical ensemble~\cite{Arcioni:2004ww}. Therefore, they are stable according to the Poincar\'{e} method,  but unstable if one considers the sign of the specific heat only.

\section{\label{sec:StabCBH} Stability of the Schwarzschild (A)dS BH in CG}

\subsection{Stability analysis: specific heat and TG}
Usual stability analysis is based on the specific heat at constant $V$. Since for (A)dS BH the volume arises as the conjugate variable of the cosmological constant, we analogously define  the specific heat at constant $\Gamma$ as
\begin{equation}\label{eq:CV}
C_\Gamma = T\;\left(\frac{\partial S}{\partial T}\right)_\Gamma
=
\frac{S\,(3\;S-2)}{2\;(3\;S-1)}
\;.
\end{equation}
Stability can be studied also by the metric $g^{\rm TG}_{\mu\nu}$,  requiring that $g^{\rm TG}_{\mu\nu}$ is a definite positive tensor. For a $2\times2$ matrix, necessary and sufficient condition is that both the determinant, $g^{\rm TG}$, and the first element, $g^{\rm TG}_{HH}$ are positive. For Schwarzschild BH in CG, one has:
\begin{equation}\label{eq:g}
g^{\rm TG} = \frac{S^6\;(1-S)^5 (3\,S-1)}{36\;\pi^4\;H^6\;(2-3\;S)^4}\;,
\end{equation}
and
\begin{equation}
g^{\rm TG}_{HH} = \frac{2\;(S-1)\;S^2\;(3\,S-4)}{H^2\;(3 S-2)^3}\,.
\end{equation}

\begin{figure}
	\centering
	\includegraphics[width=\columnwidth]{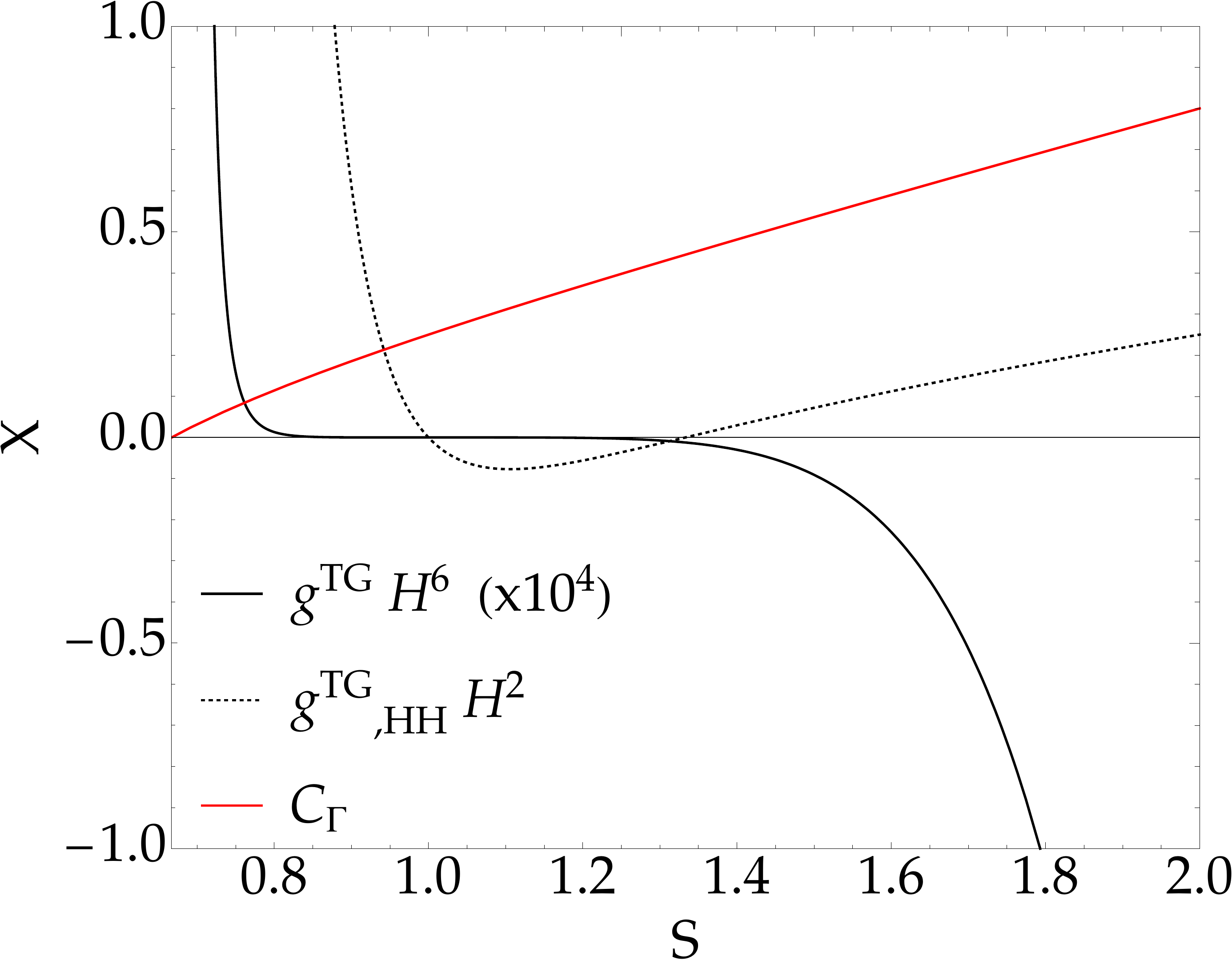}
	\caption{The specific heat (red line), $g^{\rm TG}$ (continuous black line) and $g^{\rm TG}_{,HH}$ (dotted line) for Schwarzschild BH. }
	\label{fig:gg1}
\end{figure}

Figure~\ref{fig:gg1} shows the specific heat $C_\Gamma$ (red line), $g^{\rm TG}$ (continuous black line) and $g^{\rm TG}_{,HH}$ (dotted line) for Schwarzschild BHs.

Let us recall  (see Sec.~\ref{sec:Th}) that Schwarzschild (A)dS BHs have entropy (greater)smaller than $1$ and (positive)negative $H$:
\begin{itemize}
	\item \textbf{Schwarzschild dS BHs:} both methods, the sign of $C_\Gamma$ in eq.~\eqref{eq:CV}, and the request that $g^{\rm TG}_{\mu\nu}$ is definite positive, give stable BHs for all values of the entropy;
	
	\item \textbf{Schwarzschild AdS BHs:} the two methods give opposite outcomes, as according to the sign of $C_\Gamma$, they are \textit{always} stable, while looking at $g^{\rm TG}$ and $g^{\rm TG}_{HH}$, they are \textit{never} stable.
\end{itemize}

Thus the two methods are in agreement for dS BHs in the range $2/3 < S < 1$, i.e. in a small interval of the possible values of $S$. The disagreement in other ranges of $S$ is not unexpected, though, as discussed in the Introduction~\cite{Arcioni:2004ww}. Indeed, non-extensivity plays a central role in BH thermodynamics and, in the next subsection, the Poincar\'{e} method will be applied for both dS and AdS BHs, in order to overcome these difficulties.

\bigskip

\begin{figure}
	\centering
	\includegraphics[width=\columnwidth]{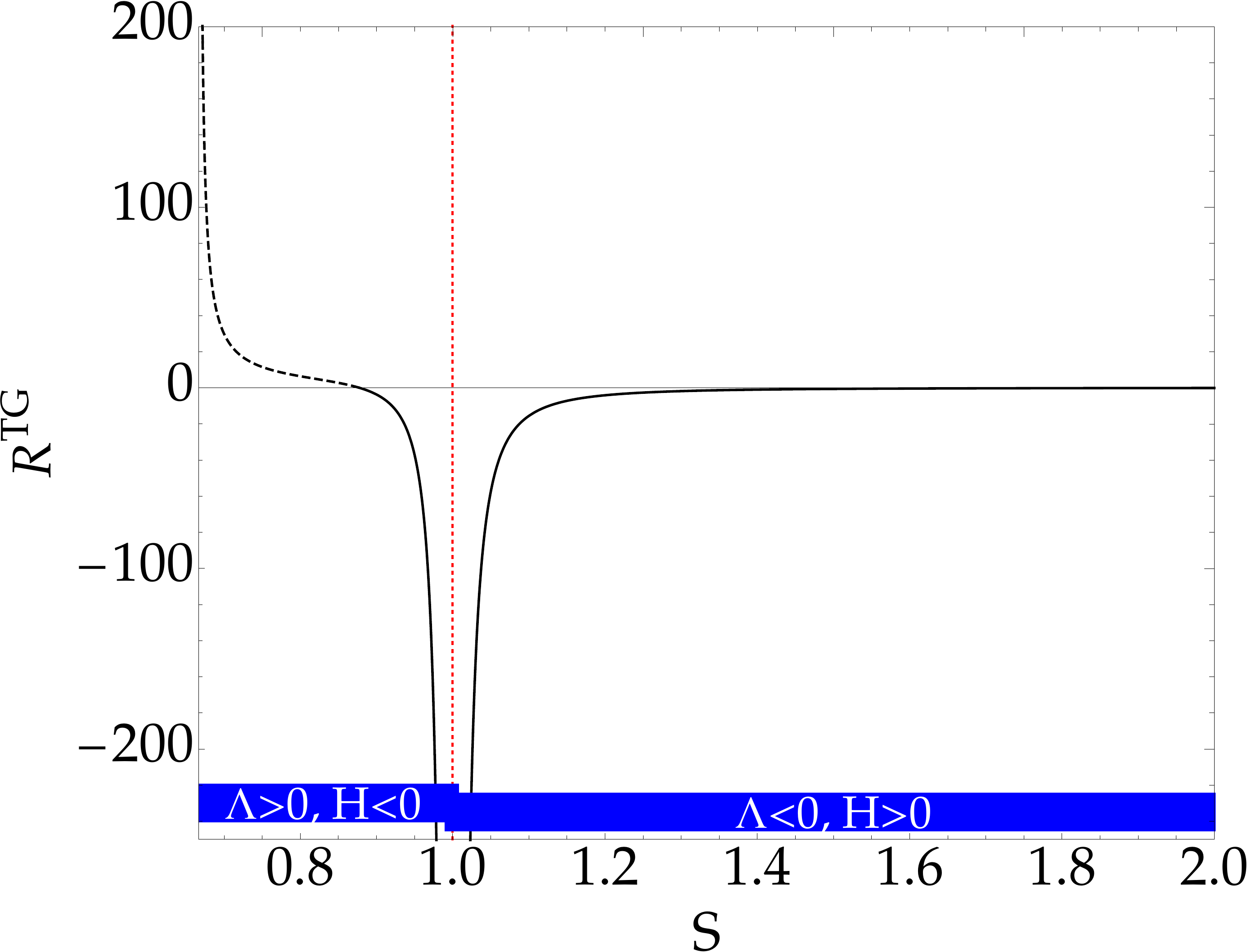}
	\caption{$R^{\rm TG}$ as a function of $S$. The dotted line refers to the range where gravitational interaction has a repulsive behavior, according to TG, hence we exclude it.}
	\label{fig:R1}
\end{figure}

The other information coming from TG are contained in the  Ruppeiner scalar curvature of Schwarzschild (A)dS BH, which turns out to be
\begin{equation}
R^{\rm TG} = \frac{5\;(4-3\;S)\;S-6}{2\;(3\;S-2)\;(S-1)^2\; (3\;S-1)^2}
\;
\end{equation}
and is shown in Fig.~\ref{fig:R1}.

\begin{itemize}
	\item \textbf{dS BHs:} for dS BHs (that have $S<1$) $R^{\rm TG}$ diverges at $S=2/3$ and $1$, as expected according to the stability analysis and defines the range of stability where both methods agree.
	
	For $S=2/3$ one gets
	\begin{equation}
	\Gamma T=0\;,
	\;\;
	H =-\frac{1}{2(9\;\pi)^2\,\Gamma}
	\;,
	\end{equation}
	\begin{equation}
	\Lambda = \frac{1}{(18\;\pi)^2\,\Gamma^2}
	\;,\qquad
	r_h = 18\;\pi\;\Gamma
	\;.
	\end{equation}
	For $S=1$ one has
	\begin{equation}
	\Gamma T=\frac{1}{(48\;\pi)^2}
	\;,\qquad
	H=\Lambda=0
	\end{equation}
	\begin{equation}
	r_h = 12\;\pi\;\Gamma
	\;.
	\end{equation}
	Moreover	$R$ changes sign at
	\begin{equation}\label{eq:S0}
	S_0 = \frac{2}{3}\left(1+\sqrt{\frac{1}{10}}\right)
	\;,
	\end{equation}
	with
	\begin{equation}
	\Gamma_0=\sqrt{\frac{35 - \sqrt{10}}{2\,\left(90\,\pi\right)^2}}\,\frac{1}{\sqrt{\Lambda_0}}
	\;,	
	\end{equation}
	\begin{equation}
	H_0=-\sqrt{\frac{35 - \sqrt{10}}{2}}\,\frac{\sqrt{\Lambda_0}}{45\;\pi}
	\;,
	\end{equation}
	\begin{equation}
	T_0 = \sqrt{\frac{5 + \sqrt{10}}{60\;\pi^2}}\,\sqrt{\Lambda_0}
	\;,
	\end{equation}
	and $\Lambda_0>0$.

	\item \textbf{AdS BHs:} ($S>1$)  $R^{\rm TG}$ diverges at $S=1$ and is always negative.
\end{itemize}

According to the usual interpretation of the sign of $R^{\rm TG}$, one might conclude that for dS BHs there is a change in the nature of the interaction of the fundamental degrees of freedom: repulsive for $2/3 < S < S_0$, attractive for $S_0 < S < 1$, where $S_0$ is given in eq.~\eqref{eq:S0}. On the other hand, for AdS BHs, $R^{\rm TG}$ always points to attractive interaction. Although, in other systems, changes of sign of $R^{\rm TG}$ have been seen~\cite{Castorina:2018gsx,Castorina:2018ayy,Castorina:2019jzw} near the critical point (and, although recent general results, on the fundamental quantum gravitational interaction, indicate a fermionic nature of such degrees of freedom \cite{BekensteinPauli2020}), we take here the view that the physical region for Schwarzschild BHs in conformal gravity requires an entropy greater than the $S_0$ in eq.~\eqref{eq:S0}, in order to have a consistent interpretation with the attractive nature of the gravitational interaction.

Moreover, $R^{\rm TG}$ diverges at $S=1$, indicating a phase transition from a dS BH ($S_0 < S < 1$) to an AdS BH ($S > 1$). Assuming that the usual ``interaction hypothesis'' of TG also holds true for non-extensive systems (i.e. $R^{TG}\propto \xi^d$ at the transition), this is a second-order phase transition, because the correlation length, $\xi$, diverges.

\subsection{Stability studied with Poincar\'{e} method}

Isolated BHs, i.e. those with $H$ as a control parameter, and $\Lambda$ and $\Xi$  fixed, are all either stable or unstable. In fact, in this case, the linear series that corresponds to $H$ is $\theta_1(H)$ and there are no vertical tangents.

\begin{figure}
	\centering
	\includegraphics[width=\columnwidth]{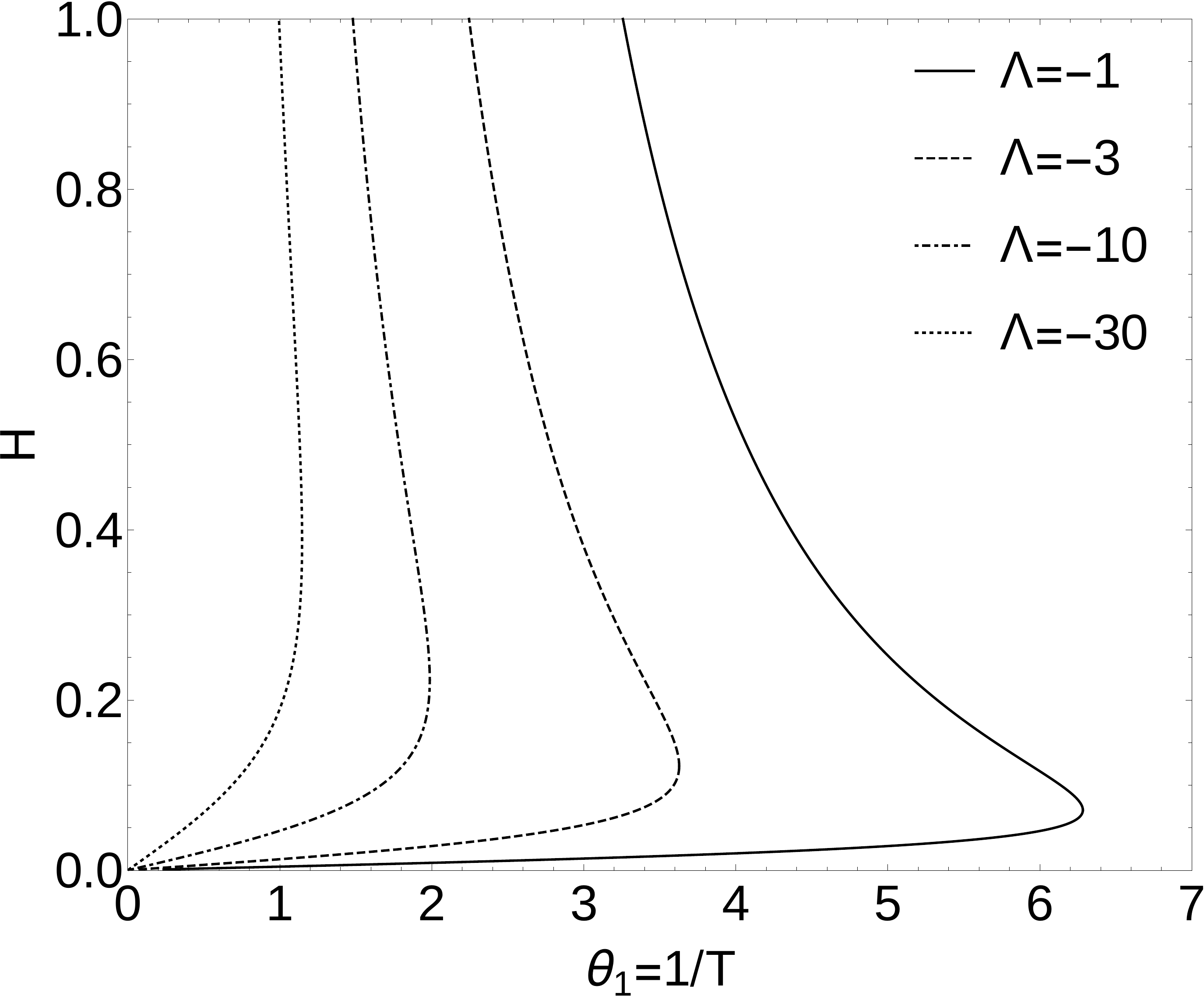}
	\caption{Plot of $H$ as a function of $\theta_1=1/T$ at $\Gamma=$const.}
	\label{fig:HvsTheta12}
\end{figure}

\begin{figure}
	\centering
	\includegraphics[width=\columnwidth]{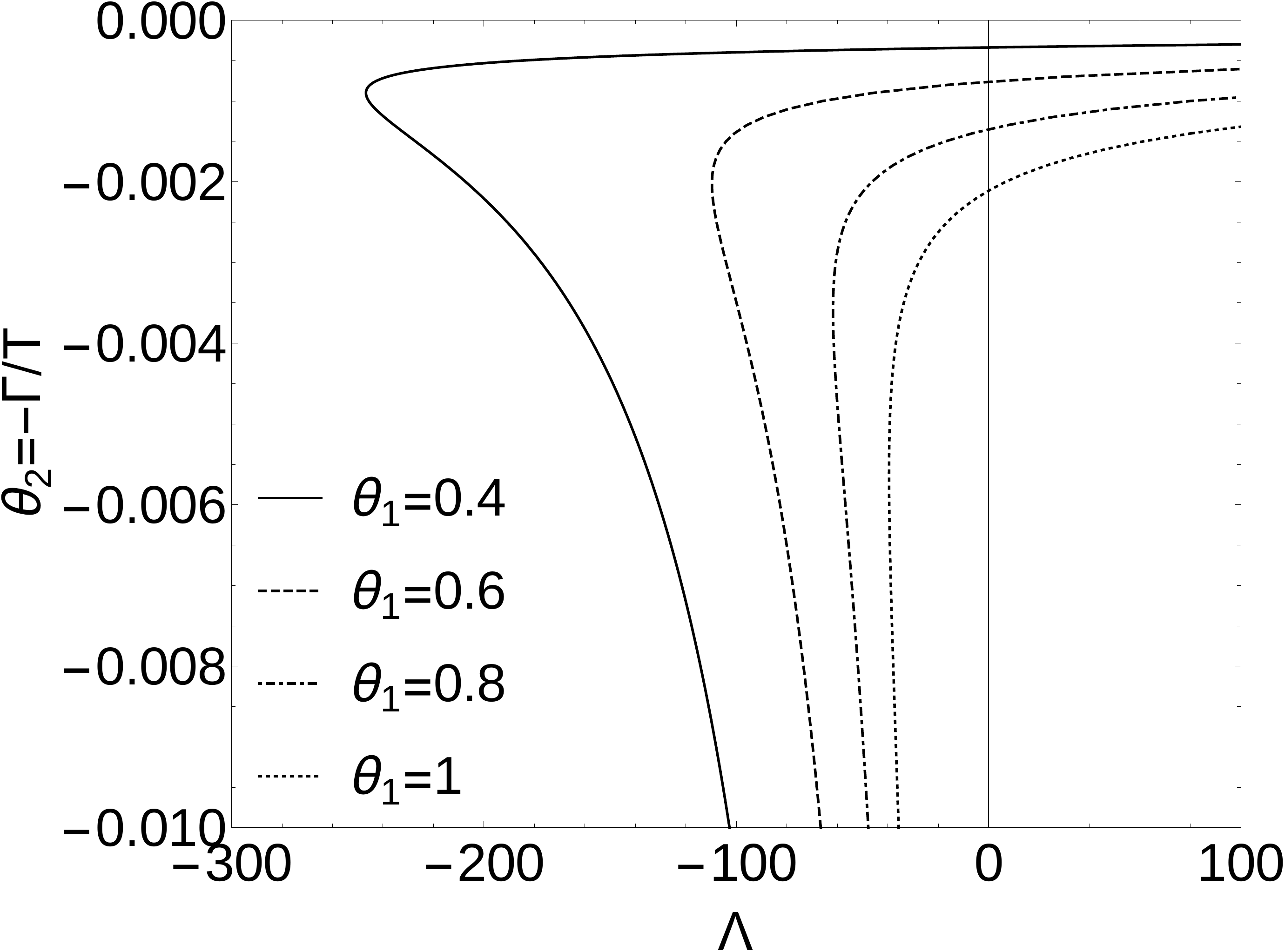}
	\caption{Plot of $\theta_2$ as a function of $P$ at fixed $\theta_1$.}
	\label{fig:Theta2vsP}
\end{figure}

Change of stability occurs only in two cases (both related to AdS BHs):
\begin{itemize}
	\item For BHs in the canonical ensemble, with enthalpy fluctuations at constant $\Lambda$ (see figure~\ref{fig:HvsTheta12}). Indeed, the linear series of $H(\theta_1)$ with fixed $\Lambda$ has a vertical tangent when $S_c=4/3$.
	
	\item For BHs with $\Lambda$ as a control parameter, and fluctuation of $\theta_2=-\Gamma/T$ at fixed $\theta_1=1/T$ (see fig.~\ref{fig:Theta2vsP}). The critical point occurs always at $S_c=4/3$.
\end{itemize}
In both cases, the critical point corresponds to a BH whose enthalpy is ($\Lambda<0$)
\begin{equation}
H_c=\frac{2\;\sqrt{-\Lambda}}{9\;\pi} \;,
\end{equation}
and with a temperature and $\Gamma$ variable given by
\begin{equation}
T_c = \frac{\sqrt{-\Lambda}}{2\;\pi}
\;,\qquad
\Gamma_c = \frac{1}{9\;\pi\sqrt{-\Lambda}}
\;.
\end{equation}
Thus they are BHs with surface gravity and radius given by
\begin{equation}
\kappa_c = \frac{1}{r_c}=\sqrt{-\Lambda}
\;.
\end{equation}
At this point stability changes from an unstable phase to a more stable one. According to the Poincar\'{e} criteria, the branch with negative slope ($S>4/3$) is  always unstable, but we can say nothing about the cases with positive slope, except that it is more stable than the previous one.

\section{\label{sec:CC}Comments and conclusions}

We studied the thermodynamics of spherically symmetric, neutral and non-rotating black holes in Conformal Gravity. We did so by means of different approaches, trying this way to test the reliability of the methods themselves, in a context still largely unexplored.

In fact, one such methods is Thermodynamic Geometry, an important area of research by itself. We showed here that it can be applied to Conformal Gravity too, because all the key thermodynamic variables are insensitive to Weyl scaling. This method focuses on the study of the concavity of the entropy, with the evaluation of the corresponding thermodynamic curvature $R^{\rm TG}$.

The other methods used here are the evaluation of the specific heat, and the Poincar\'{e} method (that is a method relating equilibrium and out-of-equilibrium thermodynamics).

The two stability analyses based on the specific heat, and on the entropy concavity, agree for dS black holes in the range of entropy, $2/3 < S < 1$, where the Thermodynamic Geometry scalar curvature, $R^{\rm TG}$, diverges at the boundary. Moreover, Thermodynamic Geometry predicts a second order phase transition, from a dS black hole to an AdS black hole, at $S=1$, but according to Poincar\'{e} criteria, black holes in Conformal Gravity are stable or in saddle point for $S < 4/3$ (and unstable for $S > 4/3$). Hence there is no definite answer on whether such transition takes place.

The non-extensivity of the entropy has a crucial role in determining the black holes stability in the parametric space, hence, for this reason, the Poincar\'{e} method should be considered more reliable than other methods.

\section*{Acknowledgements} A.I. thanks Les{\l}aw Rachwa{\l} for enlightening discussions on conformal gravity. P.C. and A.I. gladly acknowledge support from Charles University Research Center (UNCE/SCI/013).

\appendix

\section{\label{app:conti} Physical parameters}

The plot of $\Gamma T$ from eq.~\eqref{eq:S3}, i.e.
\begin{equation}\label{eq:a1}
\left(3\;S-1\right)^2 = 1+\left(12\pi\right)^2\;\Gamma T
\;,
\end{equation}
as a function of $S$ is in figure~\ref{fig:a1}. Since $\Gamma T$ is not monotonic with $S$, one defines two functions, $S_+$ and $S_-$,  as
\begin{equation}\label{eq:S4}
S_{\pm} = \frac{1\pm \sqrt{1+\left(12\pi\right)^2\Gamma T}}{3}
=
-4\;\pi\;\Psi + \frac{12\;\pi\;\Gamma}{r_\pm}
\;,
\end{equation}
where the last equality is due to the Wald formula of  eq.~\eqref{eq:S}, with $1/r_\pm$ solutions of eq.~\eqref{eq:T0}, i.e.
\begin{equation}
\frac{1}{r_{\pm}}
=
\frac{1+12\;\pi\;\Psi \pm \sqrt{1+\left(12\pi\right)^2\Gamma\;T}}{36\;\pi\;\Gamma}
\;.
\end{equation}
Therefore, the event horizon, $r_h$, is given by
\begin{equation}
r_h
=
\begin{cases}
r_- \;,& \displaystyle S<\frac{1}{3}\\\\
r_+ \;,& \displaystyle S\geq\frac{1}{3}
\end{cases}
\;,
\end{equation}
with entropy given by $S_-$ and $S_+$ in eq.~\eqref{eq:S4}, respectively. Moreover, looking at the  Schwarzschild limit $\Psi=0$, one finds that the solution $r_-$ is positive if and only if the temperature is negative (both for $\Gamma$ greater or less than zero). On the contrary, $r_+$ admits both solutions with positive and negative temperature, but $\Gamma$ is always positive.

Eq.~\eqref{eq:S2} holds for both pairs, $(r_-,S_-)$ and $(r_+,S_+)$.
Conversely, the expressions of the thermodynamic potentials $H$ and $\Lambda$ as a function of $T$, $\Gamma$ and $\Psi$ change. $\Xi$ is the same for both $(r_-,S_-)$ and $(r_+,S_+)$.

Indeed, from eq.~\eqref{eq:EH} one finds that
\begin{equation}\label{eq:PV}
\begin{split}
\Lambda_{\pm}
=&
\pm\frac{1}{9}\left(\frac{1}{2\left(6\pi\right)^2\Gamma^2}-\frac{T}{\Gamma} \right)\sqrt{1+\left(12\;\pi\right)^2\Gamma\;T}
+\\
&
+
\frac{1}{2\left(18\pi\right)^2\Gamma^2}
-
\frac{\Psi^2\left(1+4\;\pi\;\Psi\right)}{3\;\Gamma^2}
\;,
\end{split}
\end{equation}
and the  enthalpy turns out to be
\begin{equation}\label{eq:H}
\begin{split}
H_{\pm}=&
\pm \frac{2}{9} \left(T-\frac{1}{2\left(6\pi\right)^2\Gamma}\right)\;\sqrt{1+\left(12\pi\right)^2\Gamma\;T}
-\\
&
-
\frac{1}{\left(18 \pi\right)^2\;\Gamma}
-
\frac{4\;\pi\;\Psi^3}{3\;\Gamma}
\;.
\end{split}
\end{equation}
while the ``charge'' $\Xi$ is the same in the ``$-$'' and in the ``$+$'' branches and holds
\begin{equation}\label{eq:Xi}
\Xi_\pm = \Xi = -\frac{2\;\Psi\;(1+6\;\pi\;\Psi)}{3\;\Gamma}
\;.
\end{equation}
In conclusion:
\begin{itemize}
		\item BHs with entropy $0<S<1/3$ can have positive or negative $\Gamma$ and $T$, but $\Gamma T<0$.  They are described by $r_h=r_-$, have entropy $S=S_-$, $\Lambda=\Lambda_-$ and $H=H_-$. The Schwarzschild BHs in this range have positive $\Gamma$ and negative $T$;
	
	    \item BHs with entropy $1/3<S<2/3$ have positive $\Gamma$, but negative $T$. They are described by $r_h=r_+$, have entropy $S=S_+$, $\Lambda=\Lambda_+$ and $H=H_+$;

        \item BHs with entropy $S>2/3$ have positive temperature and $\Gamma>0$. They have event horizon given by $r_h=r_+$, entropy $S=S_+$, $\Lambda=\Lambda_+$ and $H=H_+$.
\end{itemize}

\section{\label{app:a2} Negative $\alpha_W$}
If $\alpha_W$ is negative,  one defines
\begin{equation}
-\frac{S}{2 \pi \alpha_W} \mapsto S
\;,
\end{equation}
in order to have a positive $S$. The relation between $S$, $\Lambda$ and $H$ now becomes
\begin{equation}
\begin{split}
\Lambda &\left[(S+1)\;S^2-32 \pi ^3 \Psi ^3\right]^2= \\
&=
12 \pi ^2 H^2 \left[(S+1)S^2+16 \pi ^2 \Psi ^2 (4 \pi  \Psi -1)\right]
\;.
\end{split}
\end{equation}
The Schwarzschild solution reads
\begin{equation}
\begin{split}
\Lambda (S+1)\;S^2
=
12 \pi ^2 H^2
\;,
\end{split}
\end{equation}
and thus, since $S>0$, Weyl gravity with $\alpha_W<0$ does not admit AdS Schwarzschild BHs.

\end{document}